\documentclass[twocolumn,showpacs,preprintnumbers,amsmath,amssymb,floatfix]{revtex4}

\usepackage{graphicx}

\usepackage{dcolumn}
\usepackage{bm}

\begin{document}


\title{Accuracy of circular polarization as a measure of spin polarization in quantum dot qubits}

\author{C. E. Pryor}
 \email{craig-pryor@uiowa.edu}
\author{M. E. Flatt\'e}
\affiliation{ Optical Science and Technology Center and Department of Physics and Astronomy, University of Iowa, Iowa City, Iowa, 52242, USA}

\date{\today}

\begin{abstract}
A quantum dot  spin LED provides a test of carrier spin injection into a qubit, as well as a means of analyzing carrier spin injection in general and local spin polarization.  
The polarization of the observed light is, however, significantly influenced by the dot geometry so the spin may be more polarized than the emitted light would naively suggest.  
We have calculated carrier polarization-dependent optical matrix elements using 8-band strain-dependent $\bf k\cdot p$ theory for InAs/GaAs self-assembled quantum dots (SAQDs) for electron and hole spin injection into a range of quantum dot sizes and shapes, and for arbitrary emission directions.  
The observed circular polarization does not depend on whether the injected spin-polarized carriers are electrons or holes, but is strongly influenced by the SAQD geometry and emission direction.  
Calculations for typical SAQD geometries with emission along $[110]$ show light that is only ~5 \%  
circularly polarized for spin states  that are 100\% 
polarized along $[110]$.  
Therefore observed polarizations\cite{Chye.prb.2002} of  ~1\%  imply a spin polarization within the dot of ~20\%.
We also find that measuring along the growth direction gives near unity conversion of spin to photon polarization, and is the least sensitive to uncertainties in SAQD geometry.

\end{abstract}


\maketitle


There are several proposals for constructing a quantum bit using a spin confined to a quantum dot\cite{Loss.pra.1998,Burkard.prb.1999,Imamoglu.prl.1999,Levy.prl.2002}.  
One method of initializing such a quantum bit  is to electrically inject spin polarized carriers 
into the quantum dot.
Towards this end, recent experiments\cite{Chye.prb.2002} have demonstrated a spin light emitting diode (spin-LED) in which spin polarized carriers are injected into and recombine within InAs/GaAs self-assembled quantum dots (SAQDs). 
The emitted light is partially circularly polarized, with the degree of polarization providing a measure of the spin in the SAQDs.
Besides its application to physical quantum bits, such a system provides information about spin transport and relaxation within the structure as a whole, which is important for the development of spin-based electronics (spintronics)\cite{Prinz.pt.1995,Wolf.science.2001}.
Spin-LEDs have also been made using quantum wells for the recombination \cite{Ohno.nature.1999, Young.apl.2002,JohnstonHalperin.prb.2002,Kohda.jjap.2001}.
The conversion of electron spin to photon polarization is filtered through the selection rules associated with the quantum well or dot.  
The selection rules for quantum wells are already controversial, and essentially nothing is known about the selection rules for circular polarization in SAQDs.
These can be complicated due to the presence of strain and uncertain geometry in SAQDs.  
In ref. \cite{Chye.prb.2002}, spin polarized electrons or holes were injected along the $[001]$ growth direction from a (Ga,Mn)As layer that was spin polarized along $[110]$.
The light  emitted along the $[110]$ direction was found to be only $\approx 1\% $ circularly polarized, suggesting a small spin polarization within the SAQD.
However, due the selection rule uncertainty, a small photon polarization does not necessarily mean that the spin polarization within the SAQD was small.

In this letter we present calculations of the circular polarization dependence of dipole recombination of spin polarized states within a self-assembled InAs/GaAs quantum dot.  
This gives a measure of the efficiency with which spin polarized SAQD states are converted into circularly polarized photons.  The calculations are done for a range of sizes, and shapes. The polarization is independent of whether the injected spin polarized carriers are electrons or holes.
We find, however, that the SAQD geometry and emission direction strongly influence the observed circular polarization which varies from 0 to $\approx 20\%$ (for 100\% polarized carriers).  
For lens a shaped SAQD there is a nonzero polarization for directions perpendicular to the growth direction only if the SAQD is elongated so as to break azimuthal symmetry.
We also show that measuring along the growth direction gives near unity conversion of spin to photon polarization, and is the least sensitive to uncertainties in SAQD geometry.

\begin{figure}
\includegraphics [width=5cm]{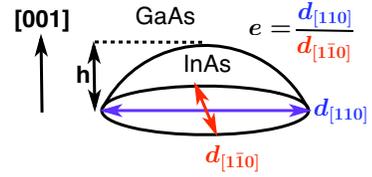}
\caption{\label{fig:1} The quantum dot geometry.    }
\end{figure}

We consider the situation in which the electron spin is polarized along a direction $d$ and the emitted light is observed along the same direction.
The emitted light is  characterized by its degree of polarization defined by
\begin{eqnarray}
P_{ d} = {   \frac {(I^+_d - I^-_d ) }  {(I^+_d + I^-_d )}  }
\end{eqnarray}
where $I^\pm_d$  is the intensity of light with $\pm$ helicity.
The InAs SAQD is taken to be an ellipsoidal cap, ellongated along $[110]$ and surrounded by GaAs  (Fig. 1).
More specifically, the geometry is determined by 
 \begin{eqnarray}
e (  {\hat x_{[110]} \cdot \vec x})^2 + 
( {\hat x_{[1\bar10]} \cdot \vec x})^2  / e+ 
( {\hat x_{001} \cdot \vec x})^2 
= r^2
 \end{eqnarray}
where $\hat x_d$ is a unit vector along the direction $d$, $e$ is the elongation, and $r$ is a scale determining the overall size of the SAQD. The ellipsoid is  sliced along a $(001)$ plane, giving the cap shown in Fig. 1.  To account for variations and uncertainties in dot geometries, we consider a range of dot shapes,  parameterized by the height $h$, the elongation $e$, and the width-to-heigth ratio $w$, 
\begin{eqnarray}
h   &=& 1.7 \rm nm, ~2.3 nm, ~2.8 nm\\
e   &=& d_{[110]}/d_{1\bar 1 0} = 1.0,~1.2,~1.4 \\
w   &=&  { (d_{[110]}+d_{1\bar 1 0})  /  h}  = 12,~16,~20
\end{eqnarray}
where $d_{[110]}$ and  $d_{1\bar 1 0}$ are the major axes along the indicated directions. 
We do not explicitly include composition gradients in the SAQD\cite{ Kegel.prl.2000}.  However, $h$ may be regarded as an effective height  after such effects are factored in.

\renewcommand{\topfraction}{.85} 
\renewcommand{\bottomfraction}{.7}
\renewcommand{\textfraction}{.15} 
\renewcommand{\floatpagefraction}{.66}
\renewcommand{\dbltopfraction}{.66} 
\renewcommand{\dblfloatpagefraction}{.66}
\setcounter{topnumber}{9} 
\setcounter{bottomnumber}{9}
\setcounter{totalnumber}{20} 
\setcounter{dbltopnumber}{9} 


\begin{figure}
\includegraphics [width=8cm]{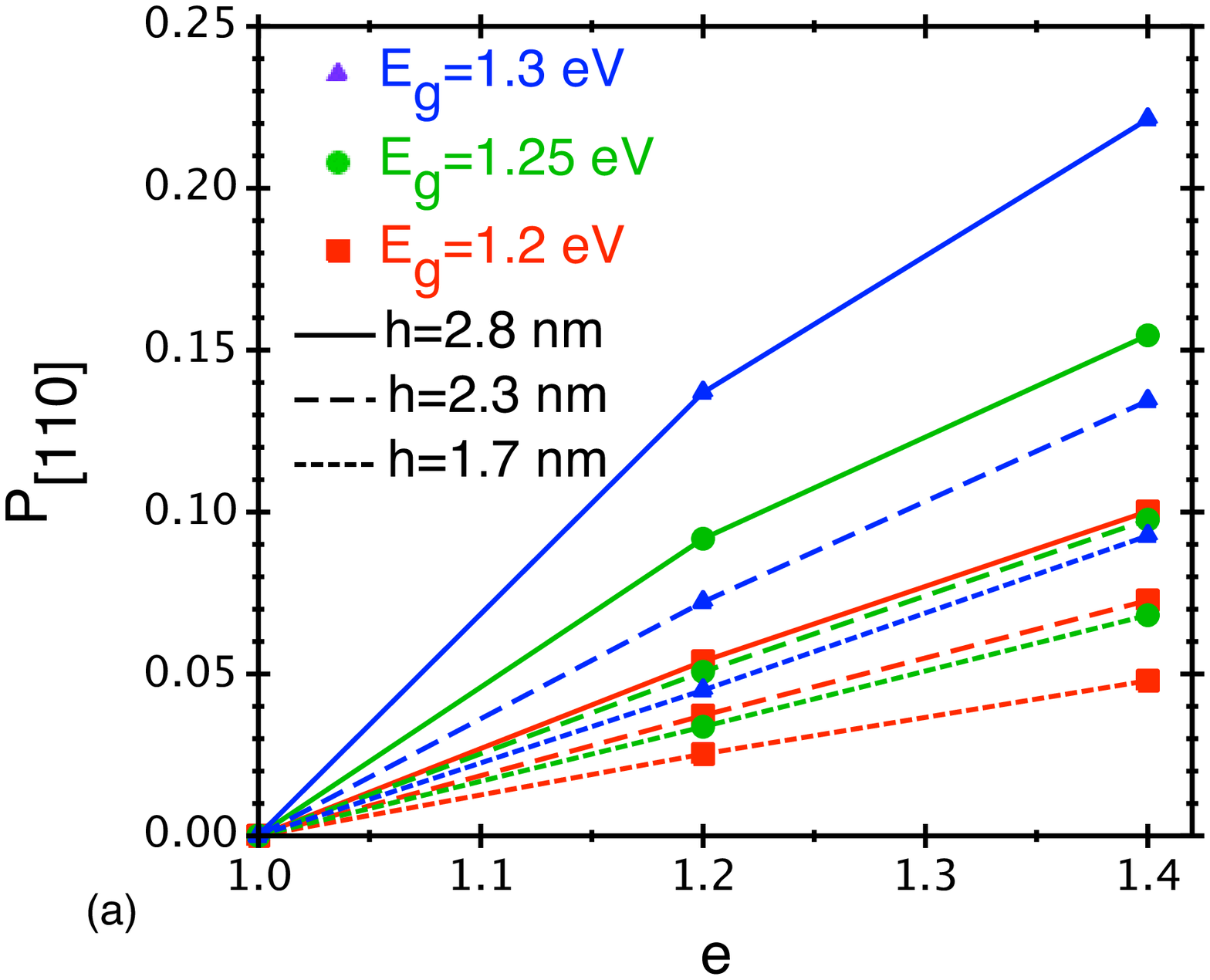}
\includegraphics [width=8cm]{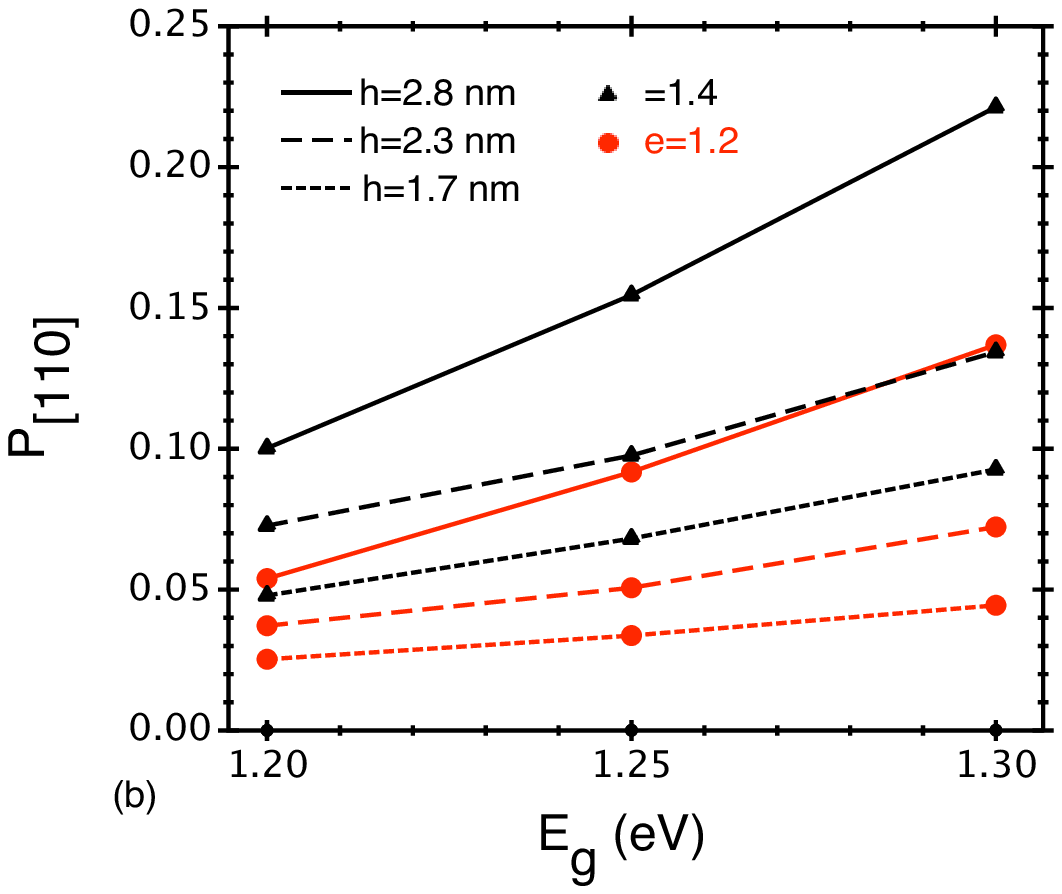}
\caption{  \label{fig:2} (a) Polarization along $[110]$, as a function of elongation.  (b) Polarization along $[110]$, as a function of bandgap.  }
\end{figure}

Ground state electron and hole wave functions were calculated using 8-band strain-dependent 
$\rm k \cdot p$ theory in the envelope approximation, by a method that has been described 
previously \cite{Pryor.prb.1997,Pryor.prb.1998}.  The grid spacing for the computation was set to the unstrained lattice constant of GaAs.
Since strain and confinement split the HH/LH degeneracy,  all levels are doubly degenerate, with states that may be denoted $| \psi  \rangle$ and $ T | \psi \rangle$, which are time-reverses of each other.  
Because the wave functions were computed with a spin-independent Hamiltonian, the state $|\psi  \rangle$ has a random spin orientation.
Spin polarized states were constructed by taking a linear combination of the states comprising the doublet, and adjusting the coefficient so as to maximize the expectation value of the pseudospin operator projected onto a direction $ d$.  
That is, we find the complex number $a$ that maximizes
\begin{eqnarray}
	{ \frac
	{\big[ 
	 \langle \psi | + a^* \langle\psi | T
	\big]
	 ~ \hat d \cdot \vec S ~
	 \big[
	  | \psi \rangle + a T |  \psi \rangle 
	  \big]
	  }
	   {(1+|a|^2)}   }.
\end{eqnarray}
The pseudospin operator in the 8-band model is given by
\begin{eqnarray}
\vec S  =
\begin{pmatrix}
\vec \sigma_{\Gamma_6}  & 0                                   & 0                                          \\
          0                                   & \vec J_{\Gamma_8}  & 0                                           \\
          0                                   & 0                                   & \vec \sigma_{\Gamma_7} \\
\end{pmatrix}
\end{eqnarray}
where $\vec \sigma$ and $\vec J$ are the spin-1/2 and spin-3/2 angular momentum operators respectively.

\begin{figure}
\includegraphics [width=5cm ]{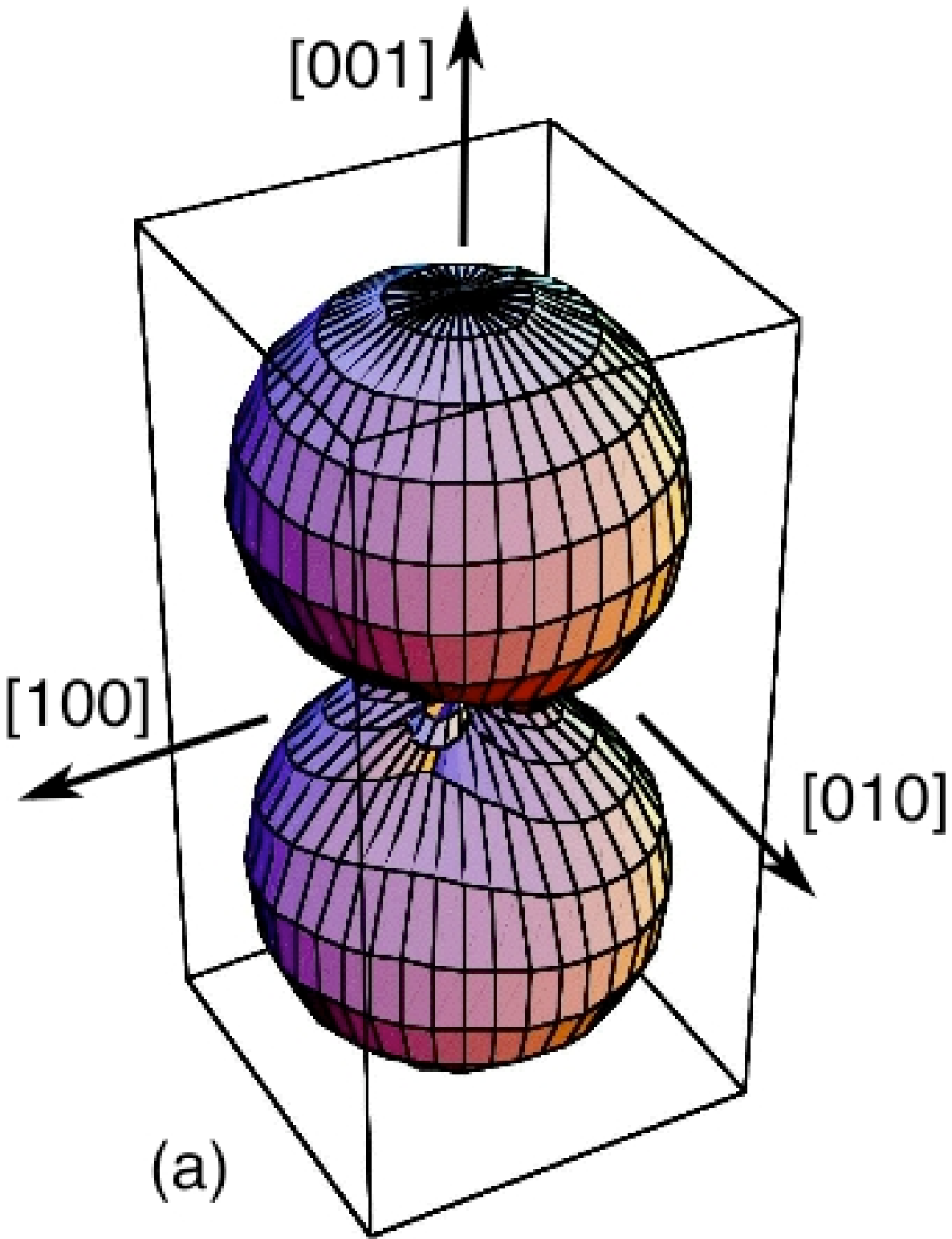}  
\includegraphics [width=8cm ]{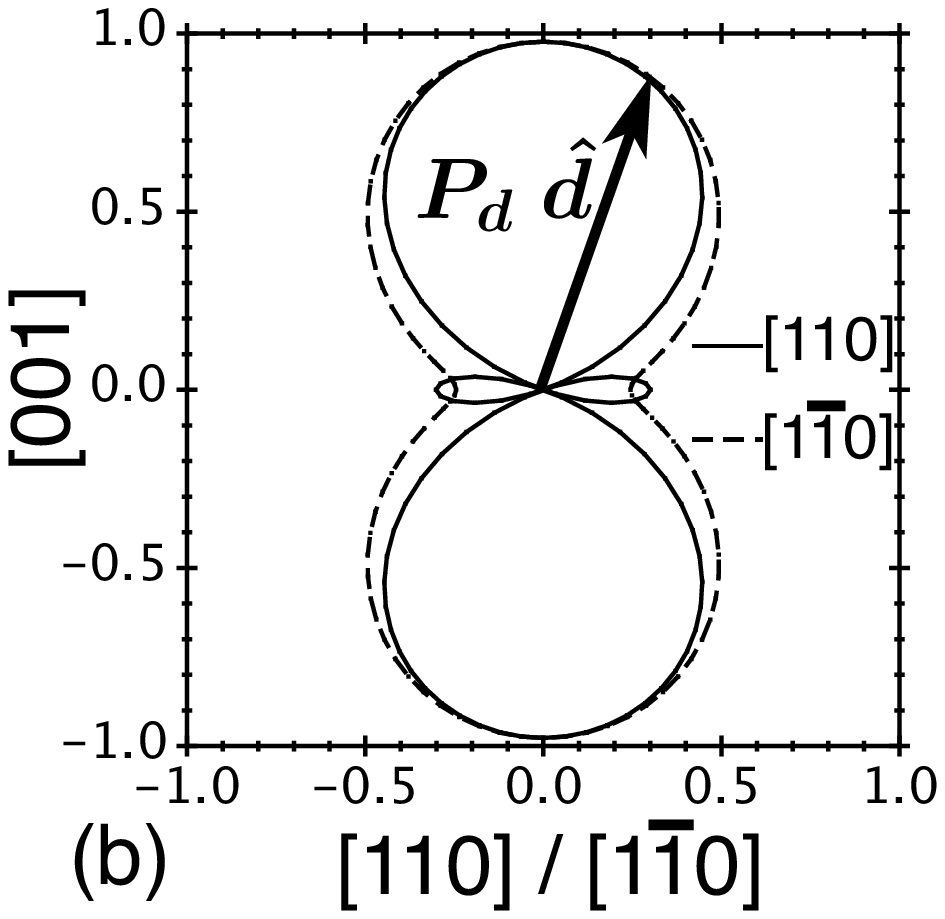}  
\includegraphics [width=8cm ]{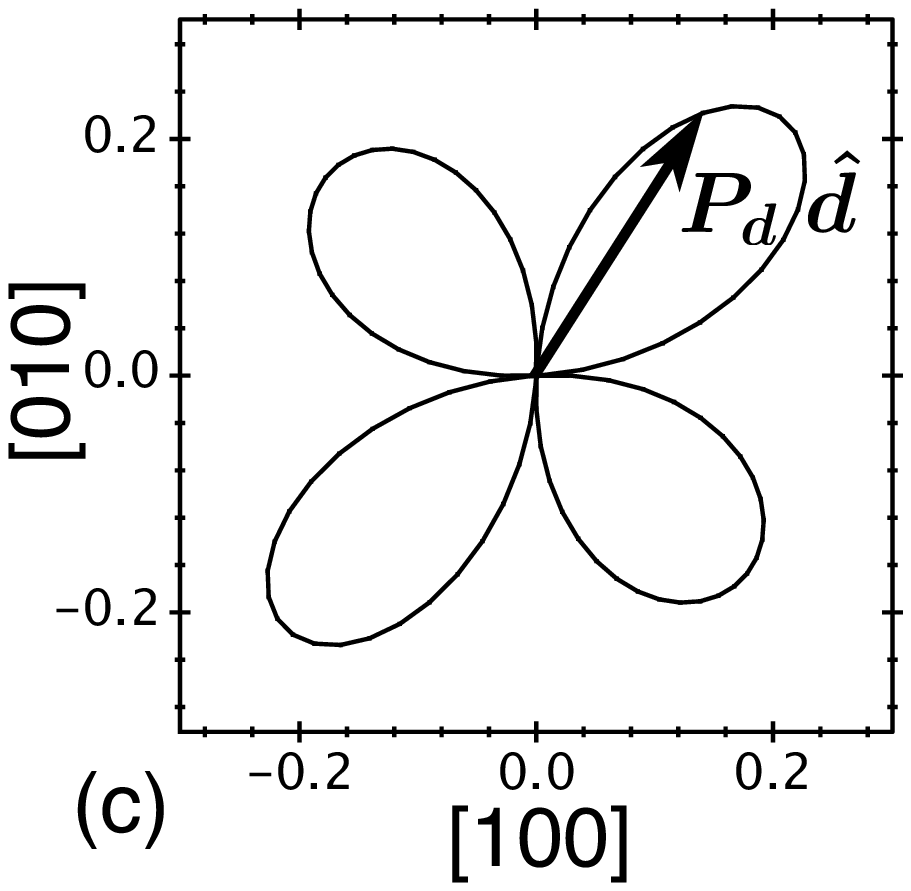}  
\caption{\label{fig:3} Polarization as a function of spin polarization and light emission direction for a SAQD with $h=2.8 \rm nm$, $e=1.4$, $E_g=1.36 eV$. The SAQD geometry was chosen to make the features more visible. (a) Surface showing 
$  P_d \hat d  = \hat d~(I^+_d - I^-_d )/(I^+_d + I^-_d ) $ (b) $  P_d \hat d $ in the $(110)$ and $(1\bar10)$ planes. (c) $  P_d  \hat d $ in the $(001)$ plane.}
\end{figure}

For spin-polarized electrons and unpolarized holes, the 
intensity for emission of circularly polarized light is given by
\begin{equation}
I^\pm_d  = | \langle       \psi_{v} |  {\bf \epsilon}^\pm_d \cdot {\bf P} | \psi_{c} \rangle |^2
                 + | \langle  \psi_{v} | T~  {\bf \epsilon}^\pm_d \cdot {\bf P} | \psi_{c} \rangle |^2
\end{equation}
where $|psi_v \rangle$ and $|psi_c \rangle$ are the conduction and valence states with spin oriented alon $\bf d$, and $\bf \epsilon^\pm_d$ is the circular polarization vector  for $\pm$ helicity  with propagation along the direction $d$.
The corresponding expression for spin polarized holes and unpolarized electrons is 
\begin{equation}
I^\pm_d  = | \langle       \psi_{v} |  {\bf \epsilon}^\pm_d \cdot {\bf P} | \psi_{c} \rangle |^2
                 + | \langle  \psi_{v} |  {\bf \epsilon}^\pm_d \cdot {\bf P} ~T| \psi_{c} \rangle |^2.
\end{equation}
However, since $T$ and $\bf P$ anticommute, equations 8 and 9 give identical results .

We first consider the case where the spins are polarized along $[110]$.
Fig. 2a shows the polarization for the $[110]$ direction, $P_{[110]}$, 
as a function of dot geomety, which is parameterized by the elongation $e$, the height $h$,  and the bandgap $E_g$.  
Rather than expressing results in terms of the width-to-heigth ratio $w$, we have linearly interpolated the final results between values of $w$ so as to obtain $P_d$ at convenient values of $E_g$. 

For axially symmetric dots ($e = 1$) the polarization is zero, but increases as the dots become more elongated.   
The polarization also increases with increasing $E_g$ with all other parameters held fixed.
Finally, for a fixed elongation and fixed band gap, shorter dots have a larger polarization.
Taken together, these indicate a general trend that the larger deviations from spherical symmetry result in larger polarization, whether comparing dimensions in the $[110]$ vs $[1\bar10]$  or $[001]$ vs $[110]$.  
	
The magnitude of the polarization is of great interest, since the experimentally observed
polarizations were only on the order of 1\%.
For the geometries considered, we find $P_{[110]} \lesssim 23\%$. 
Assuming nominal values $e = 1.2$, $h=2.3$, and $E_g = 1.25~ \rm eV$ Fig. 2 shows that for 100\% polarized carriers, the light should be 5\% circular polarized. From this value of 0.05 for the conversion efficiency, we can infer that the observed $ 1\%$ circular polarization \cite{Chye.prb.2002} was generated by carriers that were $1/0.05 \% = 20\%$ polarized.
	
Fig. 2b gives the polarization as a function of bandgap, showing the trend of increasing polarization with bandgap.  
The results of Fig. 2b disagree with the measurements\cite{Chye.prb.2002}  which show $P$ independent of $E_g$ for polarized electrons, and decreasing with increasing $E_g$ for polarized holes.
One possibility is that the polarized hole sample had growth conditions resulting in SAQDs with a size-dependent elongation.
Another likely scenario is that the dynamics of hole relaxation into the SAQD are such that spin relaxation is stronger in smaller dots.

To further examine the polarization efficiency, we performed a second series of calculations in which $P_d$  was computed as a function of the direction for a single dot geometry.
Fig. 3 shows the polarization as a function of the direction $d$ for a SAQD.  The maximum polarization is obtained along the growth direction, with $P_{[001]} \approx 0.98$.   
$P_d$ is substantially smaller along $[110]$ and $[1\bar10]$, and zero for  $[100]$ and $[010]$.  
While clearly $[001]$ is optimal, if one is restricted to the plane perpendicular to the growth direction, then $[110]$ is the best choice, while $[100]$ and $[010]$ are the worst. 
It is important to note that the photon polarization is a result of SAQD geometry and strain, and not a crystallographic effect.  $[110]$ is singled out because the SAQD is elongated in that direction.
Therefore, growth on a different substrate with different orientation will not remedy the poor spin conversion efficiency.   

Besides the larger polarization,  $[001]$ has other advantages.  For the SAQDs considered, $P_{[001]}\approx 1$ to within a few percent,  in contrast to the large variation with geometry seen for $P_{[110]} $ and $P_{[1\bar10]} $.   Thus, measuring along $[001]$ decreases  the uncertainties due to the large uncertainties in SAQD geometry.  Also, because the observed light must be collected within some solid angle, some of the emission will come from directions for which $P_d$ is smaller than the nominal direction.  From Fig. 3 we see that $[001]$ has the advantage of having less  curvature, thus decreasing effects from a non-zero collection angle.

We have shown that recombination of spin polarized carriers in InAs/GaAs SAQDs results in only modest polarization of the light emitted perpendicular to the growth direction.
The light obtains a circular polarization only for dots that are elongated so as to break azimuthal symmetry.
These results explain the small polarization seen in recent experiments \cite{Chye.prb.2002} and imply that the spin polarization of the carriers in the SAQDs in \cite{Chye.prb.2002} was $\approx 20\% $.  Measurement along $[001]$ gives the most efficient conversion of spin to photon polarization.
Finally, we have determined that measurements done along the $[001]$ direction will be less susceptible to uncertainties and variation in dot geometry, and from effects due to light being collected from a non-zero solid angle.
The implications of these features for a particular SAQD qubit will depend on the details of the implementation.
However, the anisotropy shown Fig. 3 indicates that  a SAQD differs significantly from an ideal isolated spin, and any qubit design must take this into account.

\begin{table}
\begin{ruledtabular}
\begin{tabular}{ccccc}
Parameter  &  InAs  &  GaAs \\
\hline
$E_g$ & 0.418 \rm  eV & 1.519 eV\\
$\Delta_{\rm so}$ & 0.38  eV & 0.341  eV\\
VBO & 0.085  eV & 0.0  eV \\
$\gamma_1$ & 19.67 & 6.98\\
$\gamma_2$ & 8.37 & 2.25\\
$\gamma_3$ & 9.29 & 2.88\\
$E_P$ & 22.2  eV & 22.7  eV \\
$a_c$ & -6.67   eV& -9.55  eV\\
$a_v$ & 1.67   eV& 0.95  eV\\
$b$  & -1.8   eV& -2.0  eV\\
$d$    & -3.6   eV& -5.4  eV\\
$a_{latt}$ & 0.6058 nm& 0.5653 nm\\
$C_{xxxx}$& 832.9 GPa& 1211 GPa\\
$C_{xxyy}$& 452.6 GPa& 548 GPa\\
$C_{xyxy}$& 395.9 GPa& 604 GPa\\
\end{tabular}
\end{ruledtabular}
\caption{\label{tab:table1}
Material parameters used in the calculations\cite{Vurgaftman.jap.2001}. The valence band offset (VBO) is the unstrained valence energy, determining the band alignment.  }
\end{table}

This work was  supported by the Army Research Office MURI  DAAD19-01-1-0541


\end{document}